%% file: iclr2022_conference.tex
\documentclass{article} 
\usepackage{iclr2022_conference,times}

\input{math_commands.tex}

\usepackage{hyperref}
\usepackage{url}
\usepackage{pythonhighlight}
\usepackage[ruled,vlined]{algorithm2e} 
\usepackage{graphicx}

\title{Pipeline MoE: A Flexible MoE Implementation with Pipeline Parallelism}

\author{Xin Chen, Hengheng Zhang, Xiaotao Gu, Kaifeng Bi, Lingxi Xie, Qi Tian \\
Huawei Cloud \\
\texttt{\{chenxin061, imhmhm, guxt1994, 198808xc\}@gmail.com} \\
\texttt{\{bikaifeng1, tian.qi1\}@huawei.com} \\
}

\iclrfinalcopy 
\begin{document}

\maketitle

\begin{abstract}

The Mixture of Experts (MoE) model becomes an important choice of large language models nowadays because of its scalability with sublinear computational complexity for training and inference. However, existing MoE models suffer from two critical drawbacks, 1) tremendous inner-node and inter-node communication overhead introduced by all-to-all dispatching and gathering, and 2) limited scalability for the backbone because of the bound data parallel and expert parallel to scale in the expert dimension. 
In this paper, we systematically analyze these drawbacks in terms of training efficiency in the parallel framework view and propose a novel MoE architecture called Pipeline MoE (PPMoE) to tackle them. PPMoE builds expert parallel incorporating with tensor parallel and replaces communication-intensive all-to-all dispatching and gathering with a simple tensor index slicing and inner-node all-reduce. Besides, it is convenient for PPMoE to integrate pipeline parallel to further scale the backbone due to its flexible parallel architecture. Extensive experiments show that PPMoE not only achieves a more than $1.75\times$ speed up compared to existing MoE architectures but also reaches $90\%$ throughput of its corresponding backbone model that is $20\times$ smaller. 

\end{abstract}

\section{Introduction}\label{intro}

Large language models (LLM) have shown their dominance in various natural language processing (NLP) tasks by largely pushing the limit of few-shot learning and zero-shot inference~\citep{gpt3_brown2020language, chinchila_hoffmann2022training, bloom_scao2022bloom}. Currently, it obeys a scaling law that a larger model brings better performance. While being powerful and promising, scaling up to models like GPT-3 from small models and training them are expensive and non-trivial work. 

Multiple techniques emerge to tackle the scaling problem. Tensor parallel is currently a widely applied technique that scales models up on the width dimension with a split-compute-gather manner~\citep{megatronv1_shoeybi2019megatron}. However, tensor parallel produces massive and intensive communication operations between devices, limiting its application to inner-node scenarios. Pipeline parallel, a scheme that scales models from the depth dimension, is complementary to tensor parallel and able to reach a limit of more than 500 billion parameters when incorporated with tensor parallel~\citep{megatronturing_smith2022using}. 

However, training a model of more than 100 billion parameters is time- and resource-consuming. Further scaling up is unacceptable both technically and financially. Fortunately, MoE model is an appropriate choice because it maintains scalability on the width dimension with sparsely activated experts while keeping computational complexity nearly unchanged. The MoE architecture usually cooperates with data parallel where on the device that holds one data parallel replica or rank~\citep{gshard_lepikhin2020gshard,switchtransformer_fedus2021switch,deepspeedmoe_rajbhandari2022deepspeed}, there is one or a specific number of experts on it. Thus, it is very natural and easy to scale models up by scaling the number of experts, \textit{i. e.}, number of data parallel replicas. That is to say, we can scale up MoE models almost limitlessly as long as new devices are engaged. One critical drawback of this architecture is that the computation of MoE layers follows a dispatch-compute-gather flow, which introduces two all-to-all operators on hidden embeddings, leading to a huge communication overhead that largely decreases the training efficiency.

Recent work reveals that a moderate number of experts with a thick backbone which is wider on the hidden dimension obtains better performance than a thin backbone with massive experts~\citep{jeffdean_zoph2022designing,glam_du2022glam}, which also strongly coincides with our experimental results. This indicates that the previous parallel architecture is not suitable for the new request and we need a unified scheme of scaling both expert and backbone to efficiently enlarge model capacity and capability. As depicted previously, there are existing techniques for scaling up backbones (tensor parallel and pipeline parallel) and experts (MoE). However, Adopting them together within a unified framework is non-trivial.  

In this work, we systematically analyze the drawbacks of existing MoE architecture and propose a unified framework called Pipeline MoE to efficiently and effectively scale model capacity and capability. To tackle the communication overhead of all-to-all on MoE layers, we propose a new expert parallel scheme that all experts in an MoE layer are integrated inside a node so that it can be well aligned with tensor parallel and no inter-node communication happens inside the MoE layer. Additionally, we replace the resource-consuming all-to-all communication with easy-implement tensor slicing to largely diminish the communication overhead. To better scale model capacity, we incorporate  pipeline parallel with expert parallel and tensor parallel in our Pipeline MoE, which makes our system able to scale up to a trillion-level without duplicating backbone parameters. 

Extensive experiments show that the Pipeline MoE can easily scale models to hundreds of billions of and trillions of parameters and train it $1.75\times$ faster than existing ones. Moreover, the Pipeline MoE achieves a throughput of $90\%$ its corresponding backbone model which is $20\times$ smaller.

\section{Related Work}\label{related_work}

\noindent \textbf{Scaling Up Large Language Model. }
Large language models (LLMs) have reported dominating performance of zero-shot and few-shot learning in various NLP tasks. After the introduction of transformer~\cite{transformer_vaswani2017attention}, transformer language models rapidly go across the small and medium scale~\cite{bert_devlin2018bert, gpt1_radford2018improving, gpt2_radford2019language, t5_raffel2020exploring} and step into the era of large language models with over 100 billion parameters~\cite{gpt3_brown2020language, jurassic_lieber2021jurassic, bloom_scao2022bloom, megatronturing_smith2022using, palm_chowdhery2022palm}. GPT-3~\cite{gpt3_brown2020language}, with 175 billion parameters, is the first dense LLM that pushed the limit of 100 billion parameters and achieved dramatic improvements in zero-shot and few-shot learning. Two open-source projects~\cite{bloom_scao2022bloom, opt_zhang2022opt} also published LLMs with similar sizes. Recently, the model sizes (parameter count) are further increased to an incredible 530 billion (Megatron-Turing-NLG~\cite{megatronturing_smith2022using}) and 540 billion (PaLM~\cite{palm_chowdhery2022palm}), because the scaling law~\cite{scalinglaw_henighan2020scaling} is still working. 

\noindent \textbf{Efficient Distributed Model Training. }
Scaling model training to tens of or hundreds of billion parameters is a complicated task, which requires a lot of algorithmic innovations and engineering optimization. One of the most critical challenges is that the model cannot fit into one single device or even one single node due to limited onboard memory. Fortunately, researchers and engineers proposed various techniques to jointly tackle this challenge including tensor model parallel~\cite{megatronv1_shoeybi2019megatron, sagemaker_karakus2021amazon}, pipeline model parallel~\cite{megatronv2_narayanan2021efficient, pipedreamv1_harlap2018pipedream, pipedreamv2_narayanan2019pipedream}, sequence parallel~\cite{megatronv3_korthikanti2022reducing}, checkpoint activation~\cite{checkpoint_chen2016training}, \textit{etc.} Such techniques have largely mitigated memory burden in training large language models. Our work is built upon some of these existing approaches and further increases the efficiency of MoE model training with a novel parallel architecture. 

\noindent \textbf{Mixture of Experts. }
Mixture of Experts or MoE is early introduced for machine learning applications~\cite{moe_jacobs1991adaptive, moesurvey_masoudnia2014mixture}. Recently, MoE is leveraged to enhance model capability because it is able to enlarge parameter count while keeping computational complexity nearly unchanged for both training and inference~\cite{firstmoe_shazeer2017outrageously} due to its sparse activating nature. Thanks to this fascinating property, the community achieve a huge leap in model size from a billion to a trillion parameters~\cite{gshard_lepikhin2020gshard, switchtransformer_fedus2021switch, metamoe_artetxe2021efficient}. 

Early MoE models usually applied an aggressive scaling scheme of a thin backbone with massive experts to well fit the demand of enlarging model sizes and the architecture of bound expert parallel and data parallel~\cite{gshard_lepikhin2020gshard, microsoftmoe_kim2021scalable, metamoe_artetxe2021efficient}, which, however, showed limited enhancement for model capability~\cite{deepspeedmoe_rajbhandari2022deepspeed}. This scheme and parallel strategy bring huge communication overhead due to two all-to-all operations introduced by expert parallel for each MoE layer. 
Recent works~\cite{jeffdean_zoph2022designing, glam_du2022glam} have shown that a contradictory scheme with a thick backbone and a mild expert scaling schedule results in more powerful MoE models. However, the existing parallel architecture designed for scaling experts is no longer well suited for this new strategy. On the contrary, our scheme is designed to scale both the backbone and experts, which is more flexible and configurable. 

\section{Model Architecture}\label{arch}

\subsection{Preliminary: Parallel and MoE}\label{arch_pre}

\subsubsection{Basics of Transformer}
In this paper, we take the decoder-only transformer architecture as an example to instantiate the proposed architecture. 
In a densely connected transformer, a batch of sequences with $s$ tokens is first mapped to a tensor with a shape of $b \times s \times h$ through an embedding layer (word and positional), where $b$ is the batch size, $s$ is the sequence length, and $h$ is the hidden dimension. Then intermediate states are fed through transformer blocks that are mainly composed of a self-attention module and a feed-forward network (FFN) module. Finally, the processed intermediate states are sent to the final layers to calculate output information. 
Here, we dive into a transformer block and showcase its detailed structure. The input is first normalized with a LayerNorm layer, which keeps the tensor shape unchanged. An attention module is applied to extract interaction between token embeddings. After the attention module, another LayerNorm is placed. Then an FFN is used to further extract information from input embeddings. With necessary skip connections, the processed embedding is sent to a later block. For more details, we refer readers to \cite{transformer_phuong2022formal} and \cite{megatronv1_shoeybi2019megatron}. 

\subsubsection{Data Parallel}

Data parallel (DP) is the most broadly used technique to scale up deep neural network (DNN) training in recent years. As DNN models become larger, a single GPU or device may only provide a relatively small throughput due to limited computing power and memory, while a larger batch size may be needed to ensure training stability or faster training is required. Data parallel replicates models between devices and split input data into micro-batches then feed to each replica. After all replicas finish computing (both forward and backward), model replicas on each device are synchronized by executing an all-reduce communication on gradients. 

\begin{figure}[ht]
\begin{center}
   \includegraphics[width=0.79\linewidth]{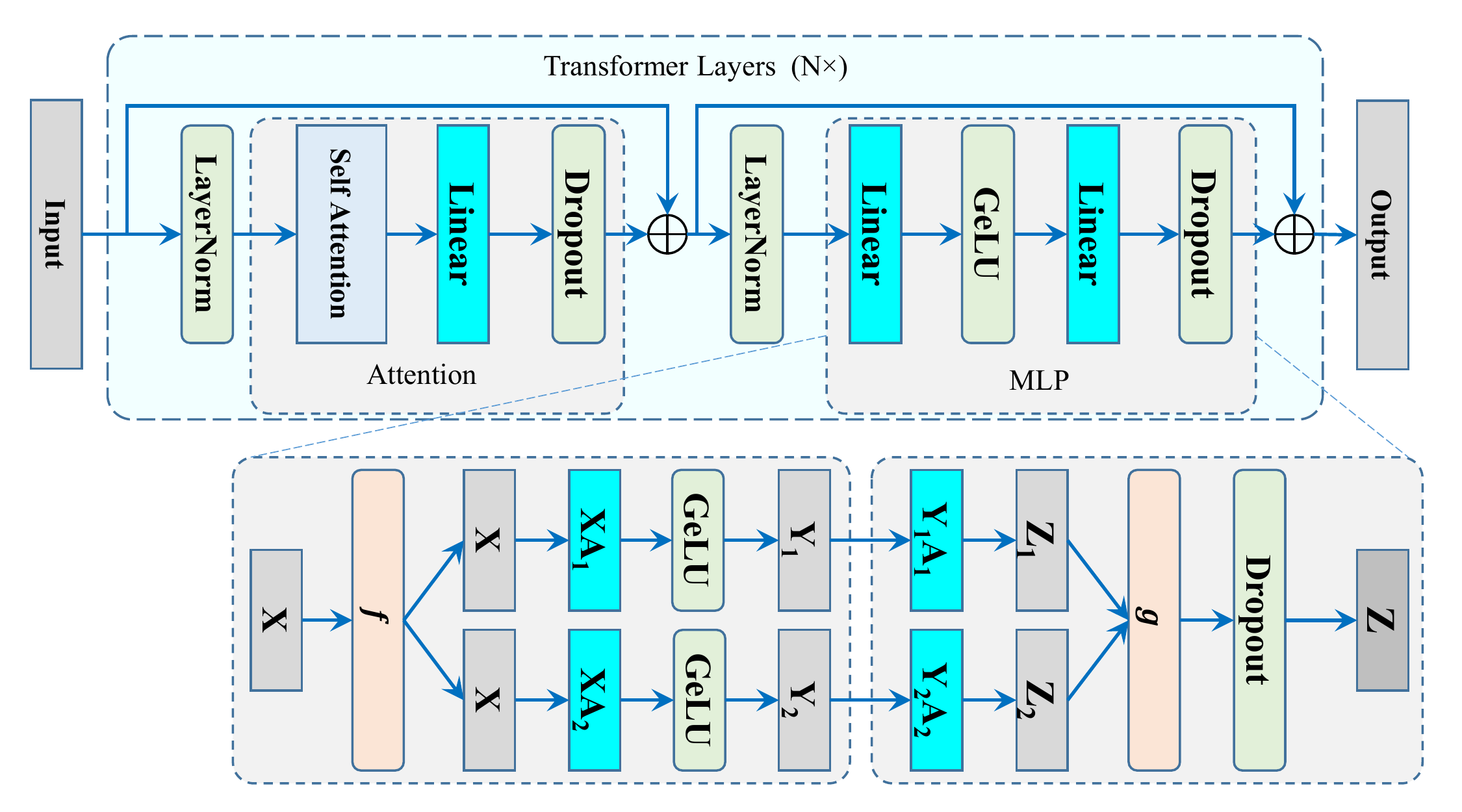}
\end{center}
\caption{Illustration of tensor parallel. A merged version of Figure~2 and Figure~3 from Megatron paper~\cite{megatronv1_shoeybi2019megatron}.}
\label{fig:tp}
\end{figure}

\subsubsection{Model Parallel}

Model parallel can be applied when models get even larger and cannot be fit into one GPU or even one node. Commonly, model parallel is composed of tensor parallel and pipeline parallel~\cite{megatronv1_shoeybi2019megatron, megatronv2_narayanan2021efficient}. 

\noindent \textbf{Tensor Parallel} Tensor parallel (TP) partitions input matrices of general matrix multiply (GEMM) into multiple sub-matrices along proper dimensions and executes smaller GEMMs inside each device. In a transformer layer, there are two groups of GEMMs that can be well-partitioned, \textit{i. e. }, the self-attention module and the FFN module. With necessary communications, a transformer layer can be well distributed into a set of devices and trained with tensor parallel. 

Here, we take the FFN module as an example to briefly introduce how tensor parallel works. An FFN module is composed of a sequence of a GEMM, a GeLU, another GEMM, and a Dropout, which can be presented as $Z = \mathrm{Dropout}(\mathrm{GeLU}(XA)B)$. Considering that $\mathrm{GeLU}(a + b) \neq \mathrm{GeLU}(a) + \mathrm{GeLU}(b)$ and $\mathrm{cat}(\mathrm{Dropout}(a), \mathrm{Dropout}(b)) \neq \mathrm{Dropout}(\mathrm{cat}(a, b))$ where $\mathrm{cat}$ represents concatenation along the proper axis, the weight matrix of the first GEMM is partitioned along its columns and the second is along its rows, as shown in Fig.~\ref{fig:tp}. Thus, for each way, all calculations are independent until Dropout, \textit{i.e.}, $Y_i = \mathrm{GeLU}(XA_i)B_i$, where $i=1,2$ in this case. In the forward pass, the Dropout operation needs to see the whole batch of input data so that the $g$ operation should be an all-reduce communication, while nothing for $f$. On the contrary, in the backward pass, gradients should be all-reduced at $f$, while nothing at $g$. In this way, we distribute parameters and calculations of GEMM equally into multiple devices with limited communication overhead (only an all-reduce for forward and another for backward). For more details of tensor parallel, we refer readers to the original paper of Megatron-LM~\cite{megatronv1_shoeybi2019megatron}. 

\begin{figure}[ht]
\begin{center}
   \includegraphics[width=0.99\linewidth]{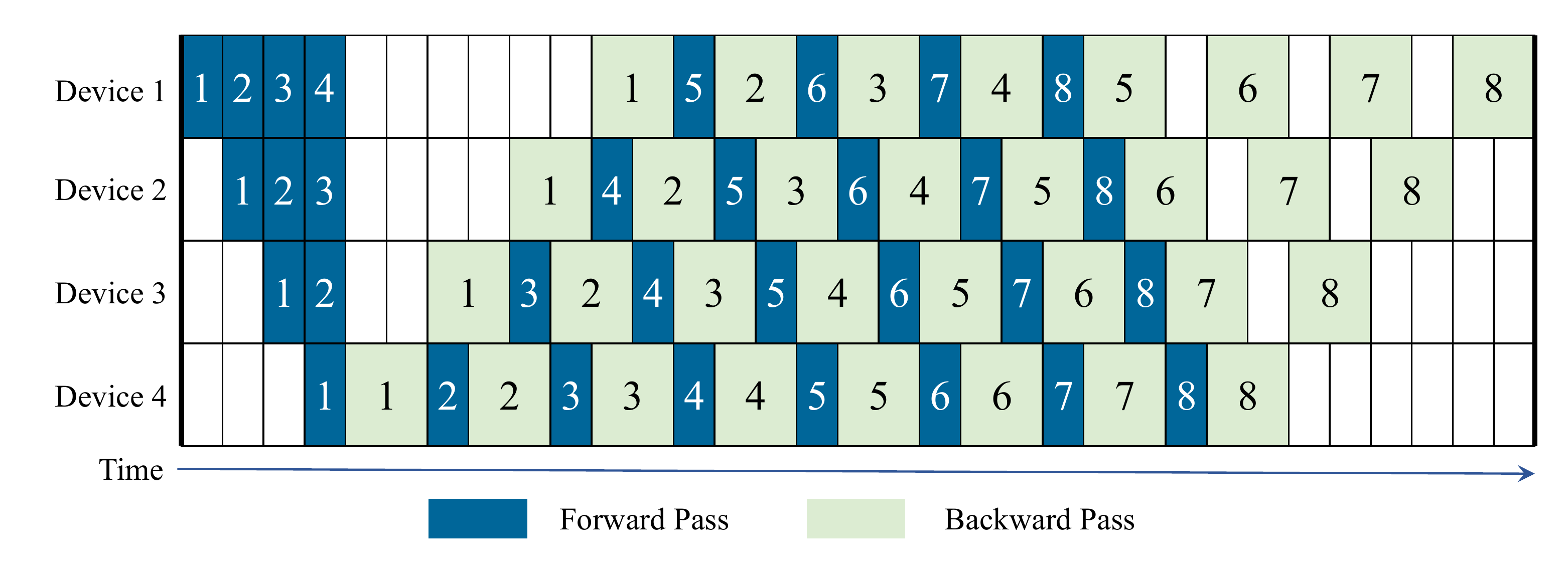}
\end{center}
\caption{Illustration of 1F1B Pipeline schedule. A re-built version of Figure~8 from PipeDream~\cite{pipedreamv1_harlap2018pipedream}.}
\label{fig:pipeline}
\end{figure}

\noindent \textbf{Pipeline Parallel} While tensor parallel can handle models with over 10 billion parameters with a single node of 8 Nvidia V100 GPUs, \textit{e.g.}, GPT-2, larger models like GPT-3 can barely fit into a single node. It is natural to split a model into multiple stages and fit each stage into different nodes, which formulate the core motivation of pipeline parallel (PP). When a former stage finishes computing, intermediate hidden states are sent to the next stage via p2p communications and continue to process in a forward pass. In backward, a latter stage finishes its backward computing, sends intermediate gradient tensors into its pre-positioned stage, and continues to execute backward computing. Such processes operate like a pipeline with a few micro-batches concurrently processed, as demonstrated in Fig.~\ref{fig:pipeline} (here we take the mostly applied 1F1B pipeline as an example). 

\begin{figure}[ht]
\begin{center}
   \includegraphics[width=0.99\linewidth]{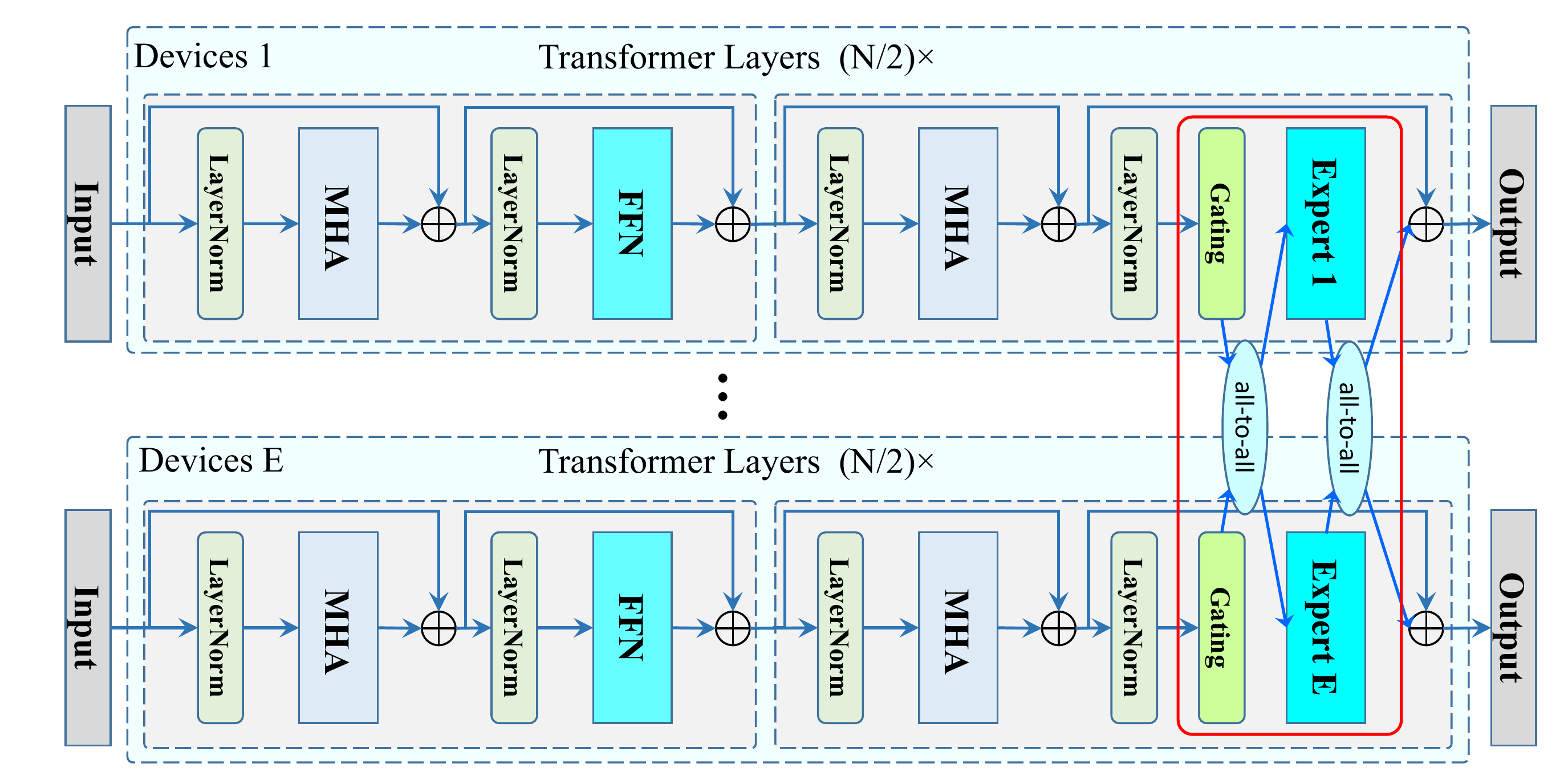}
\end{center}
\caption{Illustration of an MoE model. A re-built version of Figure~3 from GShard paper~\cite{gshard_lepikhin2020gshard}.}
\label{fig:dpmoe}
\end{figure}

\subsubsection{Mixture of Experts}\label{intro_moe}

Mixture of Experts (MoE) is first introduced in machine learning to divide a problem space into homogeneous regions and conquer them with distinct experts. Recently, MoE is leveraged into sparse model architectures to increase model capacity while keeping computational complexity unchanged. As shown in Fig.~\ref{fig:dpmoe}, an MoE system is usually bound with data parallel, and an MoE layer is built upon the FFN module and placed across multiple devices with expert parallel (EP). Inside an MoE layer, token embeddings of each data-parallel rank are firstly fed into a gating module composed of a linear layer and a softmax function, to generate scores for each token embedding that decide which expert (FFN) this token should be processed in. Then token embeddings are dispatched to their destination experts via an all-to-all communication and processed by those experts. After that, processed embeddings are combined with the pre-gating order with another all-to-all communication. 

In this way, each token embedding is automatically sharded to the expert that is well-learned to process such tokens, which implicitly strengthens the representative power of the model with negligible computational overhead. However, the price is two additional all-to-all operations for each MoE layer, which may largely slow down training and inference. Besides, a data parallel replica usually holds all layers except for MoE layers on a single GPU or a node, largely limiting the backbone size (nearly equal to the model with only one expert on each MoE layer). One way to scale up the model may be partitioning along the layer axis, but distributing experts into all devices makes it much more complicated to design the parallel scheme.

\subsection{A Systematic View on Training Efficiency of MoE}\label{analysis}

The parallel architecture of MoE is a natural choice for scaling up model size and capacity. Most existing parallel techniques including data parallel, tensor parallel, and expert parallel are adjusted into its training framework. In this subsection, we showcase the parallel architecture of existing MoE systems and analyze its advantages and drawbacks in terms of training efficiency in a systematic view. 
Two important conclusions can be drawn: \\
\indent (1) \textbf{The parallel architecture with bound DP and EP of existing MoE models is the key bottleneck of training efficiency}. \\
\indent (2) \textbf{Current parallel architecture hinders the integration of TP and PP into existing MoE models, especially TP.} \\
In the rest of this paper, we call the MoE with this typical architecture DPMoE (Data Parallel MoE) since it is firmly bound with data parallel. 

\noindent \textbf{Expert Parallel and Data Parallel}As we discussed previously, data parallel and expert parallel are bound together to guarantee that experts are evenly distributed in the training cluster, which also limits the flexibility of model architecture and training configuration. Assuming that there are $D$-way data parallel and $E$ experts, there are $E / D$ experts on each data parallel rank\footnote{$E$ is always divisible by $D$ in typical configurations. }. As introduced in subsection~\ref{intro_moe}, token embeddings need to be dispatched to each expert and return to the original data parallel rank after finishing processing on each expert, resulting in two inter-device all-to-all communication of $b\times s \times h$ data size in the forward pass. Thus, the forward time of an MoE layer is composed of gating, 1st all-to-all (\textit{1st a2a} in the equation), FFN forward, and 2nd all-to-all (\textit{2nd a2a}), as shown in Eq.~\ref{eq:forward_moe}\footnote{This equation can be well applied to the backward pass, with the gating and FFN computation time roughly doubled while communication time unchanged. }. 
\begin{equation}
    \label{eq:forward_moe}
    \centering
        t_{forward} = t_{gating} + t_{1st~a2a} + t_{FFN} + t_{2nd~a2a}
\end{equation}
Here, we only consider the situation of DP + EP and leave TP for simplicity. Since the latency of the gating module ($t_forward$) is relatively small compared to the other three items, we simply omit it here, and thus, $t_{forward} \approx t_{1st~a2a} + t_{FFN} + t_{2nd~a2a}$. According to~\cite{megatronv2_narayanan2021efficient}, the FFN module consumes $16bsh^2/E$ FLOPs on each expert. Assume that each expert is able to process F FLOPs per second, then the theoretical latency of the FFN module is $t'_{FFN} = 16bsh^2/(EF)$\footnote{This is the best case that tokens are evenly distributed in all experts. For the worst case that all tokens are processed on one expert,  $t'_{FFN} = 16bsh^2/(F)$.}. The latency of an all-to-all operation is rough $(N-1)\times(t_s + mN/(2B))$, where $t_s$ is an initial time and we omit it for simplicity since it is small compared to the other item, $B$ is the communication bandwidth, $N$ is the number of ranks involved in the all-to-all operation and is equal to $E$ in this case, and $m$ is the data count on each rank, which is equal to $2bsh/E$, where $2$ indicates that each element consumes $2$ bytes. Thus, the theoretical  latency of an all-to-all operation is roughly $t'_{a2a} = (E-1) \times bsh/B$. Combining these terms together,  we have:
\begin{equation}
    \label{eq:ratio_theory}
    \centering
       t'_{a2a}/t'_{FFN} = (E-1)EF/(16Bh).
\end{equation}
Taking the Nvidia SXM2 server with 8 V100 GPU as an example, $F=125\times 10^{12}$, $B=12.5\times 10^{9}$ for inter-node communication (InfiniBand inter-node connection in the Huawei Cloud cluster we experiment on) which is the case of PPMoE, and $h$ is generally in the range of $10^{3}\sim 10^4$. Thus, we roughly have:
\begin{equation}
    \label{eq:ratio_approx}
    \centering
       t'_{a2a}/t'_{FFN} > (E-1)E/16.
\end{equation}Thus, for normally used value of $E$, \textit{e. g.}, $64$ or $256$, $t'_{a2a} \gg t'_{FFN}$\footnote{Even for the worst case that $t'_{FFN} = 16bsh^2/(F)$, $t'_{a2a}$  is still multiple times larger than $t'_{FFN}$ since the discarded terms in Eq.~\ref{eq:ratio_theory} to Eq.~\ref{eq:ratio_approx} is larger than $1$ in practice. } and these two all-to-all operations would be a critical bottleneck of the DPMoE framework.

In practice, there are other components in the forward process, and the ratio of $t'_{a2a}/t'_{FFN}$ will be largely shrunk. We count the elapsed time of each part in a forward step of a 6.7B-to-143B DPMoE model and list numbers in Table~\ref{tab:elapsed_time}. These two all-to-all operations in MoE layers occupied $65.5\%$ of total forward time and $79.2\%$ of MoE forward time, making it still a critical bottleneck in training and inference. 
Besides, it also results in two other drawbacks: relatively low base-model capacity (small backbone) and requiring a large cluster to train the model. 

\begin{table}[h]
\caption{Components of elapsed time in a forward step. Elapsed time is in ms and the Percentage represents the proportion to the total forward time. }
\centering
\begin{tabular}{c c c c c c c }
\hline
 & Total Fwd. & MoE Fwd. & 1st all-to-all & 2nd all-to-all & Gating & Others\\
\hline
Elapsed time & 7617 & 6294 & 2566 & 2423 & 156 & 1323 \\
Percentage & $100\%$ & $82.6\%$ & $33.7\%$ & $31.8\%$ & $2.1\%$ & $17.3\%$\\
\hline
\end{tabular}
\label{tab:elapsed_time}
\end{table}

\noindent \textbf{Tensor Parallel} Tensor parallel is usually applied when a single device cannot hold the model, or one data parallel rank and its corresponding on-device experts in MoE model that roughly contains a backbone and $E / D - 1$ experts. Under such circumstances, tensor parallel is able to enlarge the backbone that can be held on a data parallel rank by almost $T\times$, where $T$ is the world size of tensor parallel and usually set to be the device count in a single node, with the price of two additional all-reduce communications of a $bsh$ tensor each self-attention and FFN module. Fortunately, these all-reduce communications are inner-node and the latency is relatively low since high-speed inner-node interconnection techniques like Nvidia NV-Link are standard configurations of GPU servers. 

Tensor parallel can be applied on all self-attention and FFN modules, or only on non-MoE blocks in the current PPMoE implementation. 
Here, we take a tensor-paralleled FFN module as an example to showcase the components of its forward latency. The forward latency of a tensor-paralleled FFN module is composed of the computational latency and the all-reduce latency:
\begin{equation}
    \label{eq:ffn_lat}
    \centering
       t_{FFN} = t_{cal} + t_{all-reduce}.
\end{equation}
As discussed previously, $t_{cal}  = 16bsh^2/(TF)$, where T is the tensor parallel world size and is usually set to be $8$, the number of GPUs inside a node. According to the NCCL document, the latency of an all-reduce operation can be formulated as $t_{all-reduce} = 2(N-1)\times(t_s + m/B))$. Ignoring the $t_s$ term and fitting to the discussed scenario, we have $t_{all-reduce} \approx 4(T-1)bsh/B$. Thus, we have:
\begin{equation}
    \label{eq:ffn_ratio}
    \centering
       t_{all-reduce}/t_{cal} = (T-1)TF/(4Bh). 
\end{equation}
Taking $F=125\times 10^{12}$, $B=300\times 10^{9}$, $T=8$ and $h=10^3$ into Eq.~\ref{eq:ffn_ratio}, $t_{all-reduce}/t_{cal} = 35 / 6 \approx 6$. 
Thus, the communication overhead of tensor parallel with inner-node all-reduce is dramatically smaller compared to expert parallel with inter-node all-to-all, which is a strong motivation for us to design the proposed framework. 

\noindent \textbf{Pipeline Parallel}Pipeline parallel is a powerful scheme to scale models to hundreds or even thousands of billions of parameters when collaborated with tensor parallel, while tensor parallel can only reach a few billion since inter-node all-reduce of large data is time-consuming. 
Currently, pipeline parallel is mostly leveraged in dense model training, \textit{e. g.}, BLOOM~\cite{bloom_scao2022bloom}, and barely applied in MoE. The reason is two folds. On one hand, existing DPMoE frameworks usually involve data parallel, expert parallel, and tensor parallel, composing a complex parallel system that requires a lot of engineering effort. Combining such a complex scheme with pipeline parallel is an even further complicated task\footnote{There may be two ways to integrate pipeline parallel into existing frameworks:  splitting the whole network (both MoE and non-MoE layers) into stages and splitting only non-MoE layers and distributing experts in the whole training cluster. The former requires a $P\times$ larger cluster to train the model since it replicates existing systems for $P$ times. The latter brings about additional communication overhead instead of resource burden because both pipeline parallel and expert parallel requires inter-node communication. Besides, the parallel scheme is much more complex since all types of parallelisms and communications are coupled and involved. }. On the other hand, with a large number of experts, the combination of data parallel, expert parallel, and tensor parallel is already able to scale models to a trillion level. However, the configuration of a small base model and massive experts may hurt the representative power of MoE models and some recent work empirically shows that a large backbone with limited experts performs better~\cite{glam_du2022glam,jeffdean_zoph2022designing}. Recall that the upper bound of tensor parallel to scale backbone is relatively low due to resource constraints of an individual node, pipeline parallel becomes a significant approach to scale MoE models. 

\begin{figure}[ht]
\begin{center}
   \includegraphics[width=0.99\linewidth]{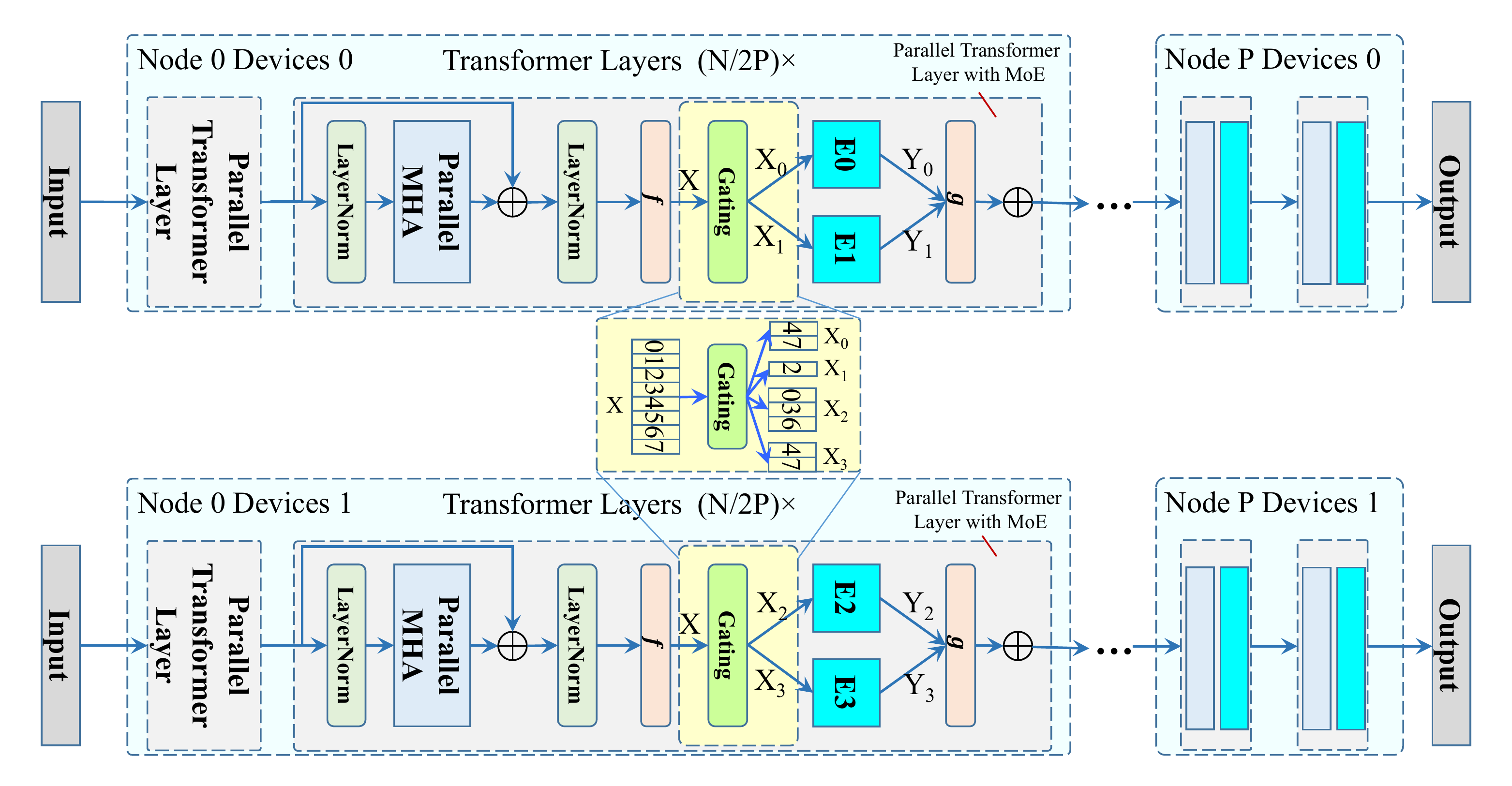}
\end{center}
\caption{Illustration of Pipeline MoE. }
\label{fig:ppmoe}
\end{figure}

\subsection{Pipeline MoE}\label{ppmoe}

To tackle drawbacks described in subsection~\ref{analysis}, we propose a novel MoE framework called Pipeline MoE (PPMoE) to efficiently improve the configuration flexibility and largely speed up its training and inference process. In this subsection, we will delve into the framework of PPMoE and show how it works. 

\subsubsection{Overview} 
Pipeline MoE decouples expert parallel from data parallel, in order to conveniently enjoy the benefits of all four parallel strategies (TP, DP, EP and PP)
Before entering into the MoE layer, hidden embeddings are synchronized by the all-reduce communication after the self-attention block so that each tensor parallel rank receives identical inputs, which is guaranteed by tensor parallel. Hidden embeddings are then fed to the gating module and generate the dispatching order. Since all inputs, parameters, and algorithms are exactly the same, the dispatching order on each rank is also identical. Then token embeddings are dispatched to their corresponding experts according to the dispatching order by an index selection operation. After processed by related experts, token embeddings are gathered by an all-reduce communication, finishing forward computing of this MoE layer. A global view of Pipeline MoE is illustrated in Fig.~\ref{fig:ppmoe}. 

\subsubsection{Expert Parallel} 
Unlike previous expert parallel of DPMoE built upon data parallel, expert parallel in our scheme is coupled with tensor parallel. On a tensor parallel group with a world size of $T$ ($T$ devices in this group), $E$ experts are evenly distributed on these $T$ devices, where there are $N$ experts on each device and $N \times T = E$. Note that there are $N$ experts on each device and we have to serialize the computation of each expert on each device. Fortunately, the computational speed of serially processing a few small tensors is nearly the same as processing a big tensor, according to our measurement with multiple experiments, indicating that there is little extra latency introduced by the serial computing of experts\footnote{This may be due to well-optimized low-level operators of PyTorch}. Such a satisfactory property guarantees a comparable computational efficiency to DPMoE on experts. 

\subsubsection{Gating and Dispatching}
The gating module of an MoE layer usually consists of a linear mapping, a softmax score function, and the gating schedule to generate dispatching orders. Our framework is compatible with existing gating schedules including top-1, top-2, \textit{etc.}  Token embeddings are then dispatched to corresponding experts with the generated dispatching order. Recall that all token embeddings and dispatching orders on each tensor parallel rank are identical and that all experts are located in the same tensor-parallel group, \textit{i. e.}, the same node, we can easily replace the communication-intensive all-to-all operation with a simple index selection operation that is well supported in most existing deep learning platforms like PyTorch~\cite{pytorch_paszke2019pytorch} or MindSpore~\cite{mindspore_huawei2022deep}. 

We instantiate this process with fig.~\ref{fig:ppmoe} as an example. Assume that we have 8 token embeddings ($X$) as input with a shape of $[8, ...]$, 2 slices in the tensor-parallel group, and 4 experts in the expert parallel group. The dispatching order is $2, 3, 1, 2, 0, 3, 2, 0$, which means that the 1st, 4th and 7th token embedding should be sent to expert 2 ($E2$), the 2nd and 6th token embedding should be processed on expert 3 ($E3$), the 3rd on expert 1 ($E1$), and the 5th and 8th on expert 0 ($E0$). Since $X$ is on all ranks of the tensor-parallel group, we can dispatch token embeddings to corresponding experts by $X_{0} = X[[4,7], ...]$, $X_{1} = X[2, ...]$, $X_{2} = X[[0, 3, 6], ...]$ and $X_{3} = X[[1, 5], ...]$. After processed by corresponding experts, output embeddings are first gathered inside each tensor parallel rank by index assignment and then collected by an all-reduce communication across the tensor parallel group. A PyTorch-like pseudo code is shown in Algorithm~\ref{ep_code}. 

\begin{algorithm}
\caption{Forward and Backward Functions in Expert Parallel}
\label{ep_code}
\begin{python}
def forward(hidden_states):
    # All-reduce gradients in the backward pass
    hidden_states = copy_to_tensor_parallel_region(hidden_states)
    logits = wg(hidden_states)
    scores = softmax(logits)
    indices, weights, l_aux = gate_function(scores)
    sliced_inputs = []
    for i in range(num_local_experts):
        sliced_inputs.append(hidden_states.index_select(indices[i]))
    expert_outputs = []
    for i, expert in enumerate(experts):
        expert_outputs.append(expert(sliced_inputs[i]))
    output_hidden_states = torch.zeros_like(hidden_states)
    for i in range(num_local_expert):
        output_hidden_states[indices[i], ...] = expert_outputs[i]
    # all-reduce output_hidden_states in the forward pass
    output_hidden_states = reduce_from_tensor_parallel_region(output_hidden_states)
    return output_hidden_states, l_aux
\end{python}
\end{algorithm}

\subsubsection{Communication Overhead}\label{comm_overhead}
In the forward pass, output embeddings with a shape of $b \times s \times h$ need to be collected via an all-reduce communication since the Dropout layer requires full access to its input data inside the data parallel rank, which is the same as tensor parallel. In the backward pass, the corresponding gradients should be synchronized before feeding into the LayerNorm layer via another all-reduce communication. Besides, parameters inside the gating module need to be synchronized in each training step via an all-reduce communication on gradients for each update (each global batch). Luckily, the communication overhead of an MoE layer for data and gradient gathering is exactly the same as tensor parallel, indicating that no extra communication overhead on data is introduced compared to tensor parallel. The only overhead is the synchronization of the linear mapping parameters in the gating module with a shape of $h \times E$. Compared to the communication overhead on data \textit{i. e.}, $2 \times b \times s \times h$ for each global batch, it is negligible since $2 \times b \times s >> E$ (4$\sim$5 orders larger). In a word, almost no extra communication overhead compared to tensor parallel is required. 

\subsubsection{Scaling with Pipeline Parallel}
Since all experts and tensor parallel partitions of the same group lie inside an individual node, models built with our proposed expert parallel scheme can be easily scaled with pipeline parallel like those dense models do~\cite{megatronv2_narayanan2021efficient}. From a layer-wise view, an MoE layer can be easily replaced by a non-MoE FFN without dramatically changing its input/output format and communication pattern. In other words, a dense model powered by tensor parallel and pipeline parallel can be seamlessly transformed into an MoE model by just replacing some of those FFNs with MoE layers build with our proposed expert parallel. Scaling MoE models with pipeline parallel in the layer axis, or equally, scaling large dense models with MoE layers in the expert axis can be easily achieved. Besides, our proposed framework is well decoupled and compatible with data parallel and could be further efficient by incorporating memory-saving techniques like ZeRO optimizer~\cite{zero_rajbhandari2020zero}. 

\subsubsection{Relationship to Prior Schemes}\label{relationship_to_prior}
DPMoE takes one micro-batch of data as input for each data parallel replica and processes these data in parallel, which composes of a global batch. On contrary, the Pipeline MoE, integrating the pipeline parallel inside, takes and processes multiple micro-batches that make up a global batch sequentially. Note that all parameters keep unchanged during processing one global batch of data and the gradient of each parameter is summed up across micro-batches for both DPMoE (by gradient all-reduce) and PPMoE (by gradient accumulation) and then updated. These two approaches could be regarded as two different patterns to span micro-batches of data in a global batch, DPMoE from a spatial view, and PPMoE from a temporal view. Thus, Pipeline MoE and previous MoE are equivalent functionally but different in parallel architectures. 

\section{Experiments}\label{exp}

In this section, we show the effectiveness of our proposed Pipeline MoE with extensive experiments. Since PPMoE is functionally equivalent to DPMoE, we only conduct a convergence verification on training/validation loss and mainly focus on speed and resource analysis. 

\subsection{Experiment Setup and Implementation Details}\label{exp_setting}
We mainly take the GPT-3 Medium with 350M parameters (24 layers, 1024 hidden size, 16 attention heads) and GPT-3 6.7B (32
layers, 4096 hidden size, 32 attention heads) as backbones or base models and scale them with $64$ experts on every other FFNs to $\sim6.7$B and $\sim143$B parameters (referred as small setting and large setting), respectively. In our experiments, we use Dense to represent backbone models, DPMoE for MoE models with previous parallel architecture, and PPMoE for MoE models with our proposed parallel architecture. The MoE gating schedule is always set to top-1, which only activates the top-ranked expert for each token. This setting guarantees that the MoE model keeps nearly the same computational complexity as its corresponding base model. A small difference between PPMoE and DPMoE is that the PPMoE abandoned the capacity limit of each expert\footnote{Although it may slightly increase computation of Pipeline MoE theoretically, we observed that the auxiliary loss term guarantees a balanced assignment of experts in most of our training. } because the max number of tokens for one expert is $bs$ while it is $Dbs$ for DPMoE. The latter makes it easier to encounter out-of-memory (OOM) errors when almost all tokens lean to choose the same expert in training. Following Gshard~\cite{gshard_lepikhin2020gshard}, we also apply an auxiliary loss term to balance expert assignment. 

The micro-batch-size of training each model is set adaptively so that the GPU and/or GPU memory utilization is above $80\%$. Following the GPT-3 paper~\cite{gpt3_brown2020language}, the sequence length is always set to be $2,048$. 
For a fair comparison, we use \textit{float16 (fp16)} for all parameters except for the MoE gating module in \textit{float32(fp32)}. We always adopt an \textit{fp16} Adam optimizer that stores parameters in \textit{fp16} and the main copy of parameters, the first and second momentums in \textit{fp32}, so that the storage for each parameter is 18 bytes. During training, ZeRO optimizer~\cite{zero_rajbhandari2020zero} is adopted in all DPMoE models with bound data parallel and expert parallel to further save memory unless unnecessary, while not in all dense models and all Pipeline MoE models. All models are trained on our private corpus composed of encyclopedia data, web data, ebook data, \textit{etc.} 

For DPMoE, we perform all the experiments with our implementation based on Megatron-LM\footnote{\url{https://github.com/NVIDIA/Megatron-LM}} v2.5 and DeepSpeed\footnote{\url{https://github.com/microsoft/DeepSpeed}} v0.5.10. For PPMoE, we build our codebase upon the implementation of Megatron-LM v2.6. It is notable that although we did not leverage DeepSpeed and ZeRO optimizer when training PPMoE models since we ignored data parallel, it is feasible to integrate those techniques into PPMoE to further scale its training with data parallel. 

\begin{figure}[t]
\begin{center}
   \includegraphics[width=0.95\linewidth]{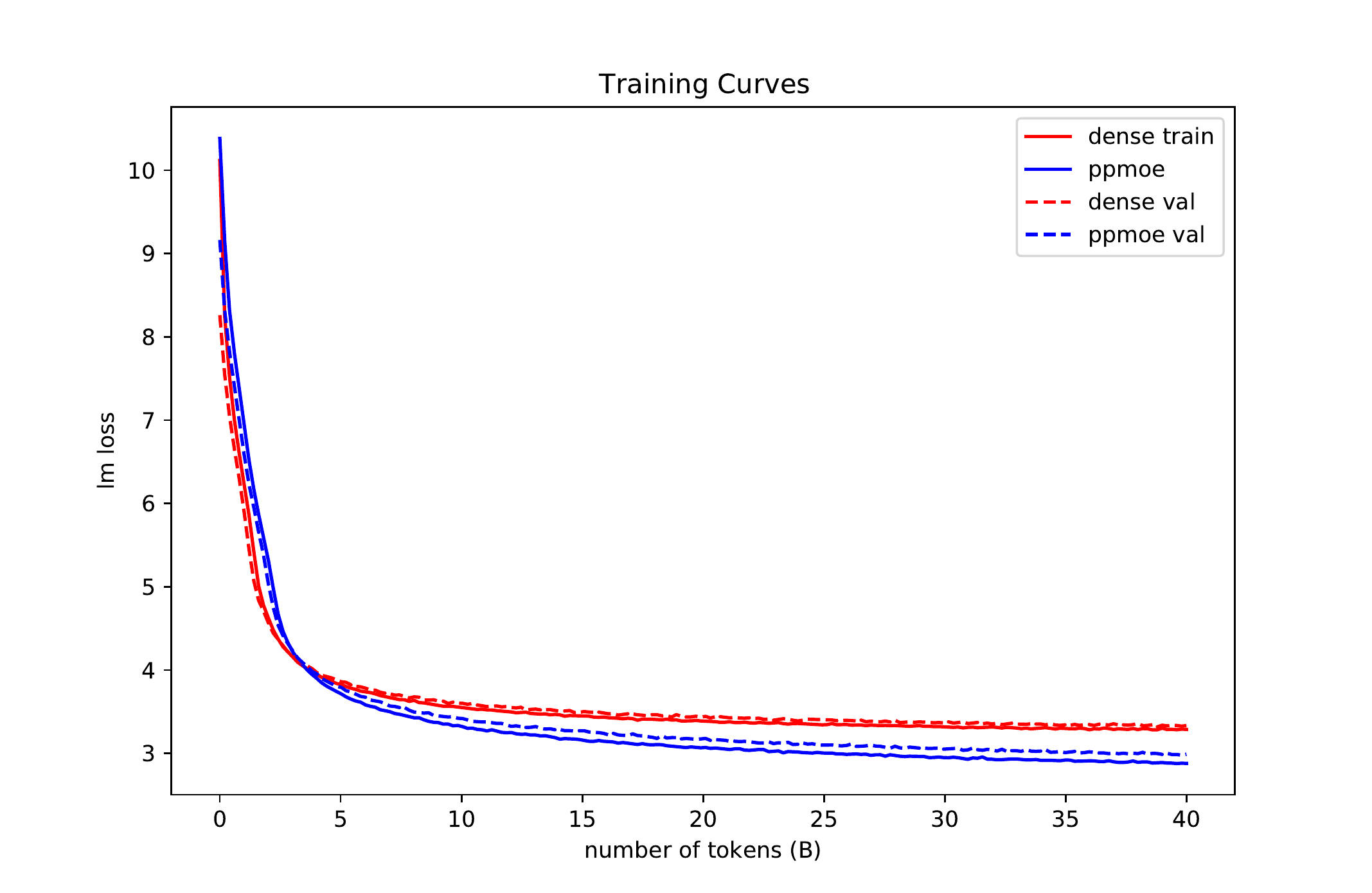}
\end{center}
\caption{Training and validation loss of 6.7B PPMoE model and its corresponding backbone model (0.3B dense) for 40B tokens. }
\label{fig:convergence}
\end{figure}

\subsection{Convergence Verification}
Firstly, we conduct experiments to verify the convergence of Pipeline MoE. We use the 6.7B PPMoE model that is scaled from a backbone of GPT-3 Medium and trains the model for 40B tokens. Both models are trained on a cluster of 4 nodes with 32 Nvidia V100 GPUs. Since the longest training process in our experiment is far from reaching an entire epoch, we just randomly sample a subset from each component of our corpus to finish the verification. Additionally, a corresponding dense backbone model with 350M  parameters (GPT-3 Medium) is also trained for 40B tokens for comparison. For the PPMoE model, the learning rate is set to be $1.2e-4$, while $3e-4$ for the backbone model, and ($\beta_1$, $\beta_2$) is set to be (0.9, 0.95) for both modes. We plot loss curves of training and validation in Fig.~\ref{fig:convergence}. As demonstrated in Fig~\ref{fig:convergence}, the training/validation loss curve of PPMoE is stable under the Dense counterpart after the initial warmup and shows a satisfactory trend of convergence. The reason that the loss of the PPMoE model is higher than its backbone at the beginning of training may lie in that the gating module needs some steps to reach a relatively stable status and so there is a small period that the training loss of the PPMoE model is higher than the backbone model. 

\begin{table}[ht]
\caption{Comparison on training speed between Dense, DPMoE, and PPMoE models. The throughput is counted as tokens per second per GPU, and the speed ratio is the ratio between the throughput of the current model and its corresponding base model. }
\centering
\begin{tabular}{c c c c c c c c c}
\hline\hline
Model & DP & TP & PP & EP & ZeRO & Training Cluster & Throughput & Speed Ratio\\
\hline
0.3B Dense   & 1   & 8 & 4 & 1   & $\times$ & 32 V100 & 3244 & - \\ 
0.3B Dense   & 4   & 8 & 1 & 1   & $\checkmark$ & 32 V100 & 4174 & - \\
0.3B Dense   & 32 & 1 & 1 & 1   & $\checkmark$ & 32 V100 & 5120 & - \\
6.7B DPMoE     & 32 & 1 & 1 & 64 & $\checkmark$  & 32 V100 & 2147 & $66.2\%$ \\
6.7B DPMoE     & 4   & 8 & 1 & 64 & $\checkmark$ & 32 V100 & 218 & $6.7\%$ \\
6.7B PPMoE & 1  & 8 & 4 & 64 & $\times$ & 32 V100 & 2708 & $81.4\%$ \\
\hline
6.7B Dense   & 1 & 8 & 16 & 1 & $\times$ & 128 V100 & 356 & - \\
6.7B Dense   & 16 & 8 & 1 & 1 & $\checkmark$ & 128 V100 & 597 & - \\
6.7B Dense   & 128 & 1 & 1 & 1 & $\checkmark$ & 128 V100 & 410 & - \\
143B DPMoE    & 256 & 1 & 1 & 64 & $\checkmark$& 256 V100 & 93 & $26.1\%$ \\
143B DPMoE    & 128 & 2 & 1 & 64 & $\checkmark$& 256 V100 & 183 & $51.4\%$ \\
143B DPMoE    & 32 & 8 & 1 & 64 & $\checkmark$ & 256 V100 & 63 & $17.7\%$ \\
143B PPMoE & 1 & 8 & 16 & 64 & $\times$ & 128 V100 & 323 & $90.7\%$ \\
\hline\hline
\end{tabular}
\label{tab:training_speed}
\end{table}

\subsection{Training Speed Analysis}
We conduct extensive experiments to verify the effectiveness of the proposed method in terms of training throughput (number of tokens processed per second per device). For Dense models or backbone models, we report performance on three settings of parallel schemes: the combination of tensor parallel and pipeline parallel, the combination of tensor parallel and data parallel, and data parallel only. For DPMoE models, we test two settings of parallel schemes: the combination of data parallel and expert parallel, and the combination of data parallel, tensor parallel, and expert parallel. For PPMoE models, we only test one setting and just ignore data parallel since data parallel only scales the batch size of training. Most models with data parallel are trained with ZeRO optimizer to reduce memory overhead. It is notable that the 143B DPMoE model is not able to fit into 16 nodes (128 V100 GPUs) without involving tensor parallel, so we increase the cluster to 256 V100 GPUs to make it possible. 

Experimental results are listed in Table~\ref{tab:training_speed}. For Dense models, pipeline parallel is a critical factor that largely slows down training, which is the reason that state-of-the-art dense models like BLOOM omits pipeline parallel when models are not extremely large and can be fit into a single node. For a fair comparison, we take the slowest ones as baselines to calculate the speed ratio. For DPMoE models of both the small setting and the large setting, the DP + EP setting achieves higher throughputS ($66.2\%$ and $26.1\%$, respectively) than the DP + TP + EP setting with TP = 8 ($6.7\%$ and $17.7\%$, respectively). 
This is because, with a large TP size, the communication overhead is relatively heavy. The MoE layer in DPMoE introduces two additional all-to-all operations in the forward pass, further intensifying the communication burden. 
For the large setting, a better balance with DP + TP + EP with TP = 2 achieves a higher throughput of $51.4\%$ baseline than the other settings. The reason that the one with TP is better than the one without TP is that the time complexity of ring-style all-to-all communication is proportional to the number of processes and a larger world size of expert parallel brings longer communication time. 
The above results indicate that it is non-trivial to efficiently scale with DPMoE since there are many factors that can largely influence the performance. 

One can easily discover that PPMoE models consistently achieve high throughput for both small and large settings. Specifically, for the small setting, PPMoE obtains a $24.6\%$ improvement over DPMoE, while for the large setting, the improvement goes further to $76.5\%$. The consistent improvements clearly verify the effectiveness of our proposed Pipeline MoE architecture. We notice that the improvement in the large setting is much more obvious than it in the small setting. This is because the relative cost of pipeline parallel is heavier on small models than on large ones, which also coincides with the phenomenon on Dense models. 

\begin{table}[ht]
\caption{Components of elapsed time in a forward step of Pipeline MoE. Elapsed times are listed in the second line and in ms, and the percentage to total forward time is in the third line. Numbers in \textcolor{blue}{blue} and \textcolor{cyan}{cyan} indicate that they belong to the same process and numbers in \textbf{bold} is the total elapsed time for the process. }
\centering
\begin{tabular}{c c c c c c c c}
\hline
Total Fwd. & MoE Fwd. & Gating & Exp. Calc. & MoE A.-R. & FFN Fwd. & FFN A.-R.\\
\hline
\textbf{6257}& \textbf{\textcolor{blue}{2393}} & \textcolor{blue}{491} & \textcolor{blue}{437} & \textcolor{blue}{1294} & \textbf{\textcolor{cyan}{1714 }}  & \textcolor{cyan}{1176}\\
$100\%$ & \textbf{\textcolor{blue}{$38.2\%$}} & \textcolor{blue}{$7.8\%$} & \textcolor{blue}{$7.0\%$} & \textcolor{blue}{ $20.7\%$} & \textbf{\textcolor{cyan}{$27.4\%$}}  & \textcolor{cyan}{$18.8\%$}\\
\hline
\end{tabular}
\label{tab:elapsed_time_ppmoe}
\end{table}

\subsection{Components of Elapsed Time for Pipeline MoE}
To better depict the resource consumption of the training process, we count components of elapsed time for a single forward step of a small setting PPMoE model and list elapsed time for dominating operations in Table~\ref{tab:elapsed_time_ppmoe}. Compared to DPMoE which spends $82.6\%$ of forward time in MoE forward, PPMoE decreases it to $38.2\%$. In DPMoE, main communication (two all-to-all operators) occupies $65.5\%$ of total forward time and $79.2\%$ of MoE forward time, while PPMoE reduces them to $20.7\%$ and $54.1\%$. It is notable that the MoE all-reduce time is very close to FFN all-reduce time (only $1.9\%$ of total forward time difference), which coincides with our functional analysis in~\ref{comm_overhead} and evidently verifies the effectiveness of eliminating all-to-all communication in MoE. 

Delving deeper into the MoE forward time and FFN forward time, we can find that the inner-node all-reduce communication occupied a large proportion of forward time ($54.0\%$ and $68.6\%$, respectively). Thus, there is more room for further speeding up training if a faster all-reduce scheme is adopted.

\section{Conclusion and Discussion}

In this work, we introduce PPMoE, a novel MoE framework that efficiently solves two critical drawbacks of DPMoE. By replacing communication-intensive all-to-all dispatching and gathering of DPMoE with a simple tensor slicing and an inner-node all-reduce, PPMoE largely eases the communication burden of MoE models. By incorporating with pipeline parallel, PPMoE enhances the scalability of the backbone of MoE models. Experimental results have demonstrated the effectiveness of the proposed framework. 

Our work paves a new step towards decoupling expert parallel from data parallel and eliminating the communication-intensive all-to-all operations from expert parallel. We expect that in the future with more powerful computational resources and parallel computing techniques, MoE models with reduced communication burden and increased configuration flexibility plays a more significant role in building large language models and their variants for better training/validation efficiency and representative capability. 


\bibliography{iclr2022_conference}
\bibliographystyle{iclr2022_conference}


\end{document}

%% file: math_commands.tex
\usepackage{amsmath,amsfonts,bm}

\def\eqref#1{equation~\ref{#1}}

\def\1{\bm{1}}

\DeclareMathAlphabet{\mathsfit}{\encodingdefault}{\sfdefault}{m}{sl}
\SetMathAlphabet{\mathsfit}{bold}{\encodingdefault}{\sfdefault}{bx}{n}

